# Power-Based Real-Time Respiration Monitoring Using FMCW Radar

Sherif Abdulatif*, Fady Aziz*, Pelin Altiner, Bernhard Kleiner, Urs Schneider
Department of Biomechatronic System, Fraunhofer Institute for Manufacturing Engineering and Automation IPA
Email: {sherif.abdulatif, fady.aziz, bernhard.kleiner, urs.schneider}@ipa.fraunhofer.de
* These authors contributed to this work equally.

*Abstract*—Non-contact vital sign detection is a required application nowadays in many fields as patient monitoring and static human detection. Within the last decade, radar has been introduced as a smart and convenient sensor for non-contact respiration monitoring. Radar sensors are considered suitable for such application for its capability to work through obstacles and in harsh environmental conditions. FMCW radar has been introduced as a powerful tool in this field for its capability of detecting both the breathing target position and his chest micro-motions induced due to breathing. Most of the presented techniques for using the radar for respiration detection is based on bandpass filtering or wavelet transforms on the required harmonics in either the range or Doppler dimension. However, both techniques affect the real-time capability of the monitoring and work on limited distances and aspect angles. A recognizable fluctuation effect is observed in the received range spectrum overtime due to respiration chest movements. The proposed technique in this paper is based on detecting and processing the power changes in real-time over different aspect angles and distances. Two radar modules working on different carrier frequency bands, bandwidths and output power levels were tested and compared.

## I. Introduction

Human detection is a challenging research aspect that has been addressed in many research fields nowadays. Radar sensors proved to be feasible for human detection applications for its capability of detecting the velocity and range of a target. Accordingly, radar can be a used as a convenient sensor for detecting limbs motion distribution of a moving human subject [1]. Moreover, radar can be useful for detecting static human targets based on either the human subject Radar Cross Section (RCS) [2] or vital sign (respiration and heart beats) detection based on micro-motions induced in the chest of a static human [3]. Accordingly, radar is investigated as a convenient non-contact vital sign detection sensor for patient monitoring compared to the traditional systems [4].

The radar capability of achieving the respiration monitoring with light weight, lower power and better accuracy allows the possibility of many applications in which human safety is highly required. One example is vital sign monitoring as an assistive tool for the car driver for fatigue detection [5]. The radar has also been proved to be feasible for some applications like rescue missions that require detecting presence of a living human victim to provide critical health support. In such application, a high output power radar that can detect chest micro-motions on far distances and different aspect angles is required. Three types of radar were used for the non-contact vital sign monitoring which are the Continuous Wave (CW) radar, Ultra-Wide-Band (UWB) pulse radar and Frequency Modulated Continuous Wave (FMCW) radar.

**CW Radar** technique is based on detecting the reflected frequency echoes due to the chest wall motion. These frequencies can be extracted through signal processing analysis such as frequency mixing and filtering. Bandpass filters can be directly applied for the harmonic separation induced due to both the respiration and heart rates [6]. Another techniques were used to extract heart beats with better accuracy by evaluating the wavelet coefficients in frequency band based on using wavelet transform [7]. CW radar offers simplicity in design and operation, thus they have relatively low power consumption. However, the target range data cannot be detected by CW radar and the vital sign signals is not differentiable from other received frequency echoes.

**UWB Pulse radar** has been recently introduced for non-contact respiration monitoring. The basic operation is sending a train of pulses towards the target and then the received signal can be visualized in the frequency domain which is not directly correlated with chest wall motion. To get an accurate estimation of vital signs, the delay profile in time domain is extracted using IFFT which is correlated to absolute amplitude of the chest movements. The UWB pulse radar can be used for higher spatial resolution, however, there are constraints on the pulse width and the radar peak signal intensity [8].

**FMCW Radar** has both advantages of identifying the target range available in the UWB radar and CW radar robustness. These properties introduce the FMCW radar to be more convenient for the vital sign detection task especially in multi-targets scenario. Moreover, FMCW power consumption is low with relatively small compatible sizes. Basic technique for vital signs extraction relies on the use of band pass filters to extract harmonics related to chest micro-motions, these ways have a maximum distance of 2 m for detection [9]. More advanced techniques based on wavelet transforms and eigenvalue decomposition is used; however, such techniques are complex and affects the real-time capability [10].

In this work, we propose a real-time non-complex technique for respiration monitoring at long distances and different aspect angles using FMCW radar. In the proposed technique, the power fluctuations in the range profile living target peak are observed to be correlated with the target respiration behavior. This technique provides a simple platform for detecting chest micro-motions at different aspect angles and distances above 4 m. This can help in victim detection applications, in addition to general safety applications where static human target detection is required.

## II. POWER-BASED DETECTION METHOD

### A. FMCW Operation

The operation of FMCW is based on sending continuous modulated chirps. During the chirp time, the chirp frequency increases linearly within the operation bandwidth. The received signal is down-converted with the radar carrier frequency for target range and velocity estimation. The velocity information is extracted from the frequency shifts due to the Doppler effect of the moving target. While the time delay can be used to get the target range information. The target range information is extracted through Fast Fourier Transform (FFT) that is directly applied on each received chirp. This transformation yields consecutive range profiles in which the detected targets appear as different peaks at different frequency shifts. Each shift represents a certain target range and the power values are correlated with the target RCS.

Varying environmental conditions and measuring at long distances may induce unwanted targets that appear as clutters in the analyzed range profiles by the radar. Thus, an adaptive Constant False Alarm Rate (CFAR) technique is used in removing the unwanted noises and clutter effects on real-time basis [11]. The idea of operation is based on preserving a constant rate for the detection of false targets in the case of variable clutter and noise effects. In this paper, the Cell-Averaging (CA-) CFAR is used on range spectrum for its simplicity. The proposed technique is based on using a sliding window over the range profiles in which the average power of the window cells is computed. Within one sliding window analysis, a possible peak is expected. Thus, the power averaging is computed without taking the middle cells into consideration. Finally, a threshold is constructed to filter out the real peaks from noise and clutters. Depicted in Fig. 1, an extracted target in range spectrum based on CFAR thresholding.

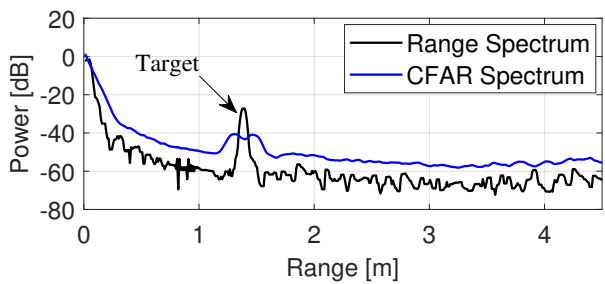

Fig. 1: Range profiles with peak at the detected target.

### B. Respiration in Power Domain

"Inhalation and exhalation" the two main respiration activities, induce an observable chest movements. While person is sitting in front of the radar, it is seen that aforementioned chest motions cause changes in the range spectrum. The changes are particularly observed as range shifts in the target peak due to the chest micro-motions, but to detect such motions an accurate radar is needed. Moreover, fluctuations in the power value of such peaks is observed in the signal overtime. Such effect can be observed by concatenating consecutive range profiles overtime. As shown in Fig. 2, a 2-D range waterfall is used to analyse power of range spectrum over time. In this plotting, the third dimension representing power (colour) of detected target (single line) is observed to change during respiration instants.

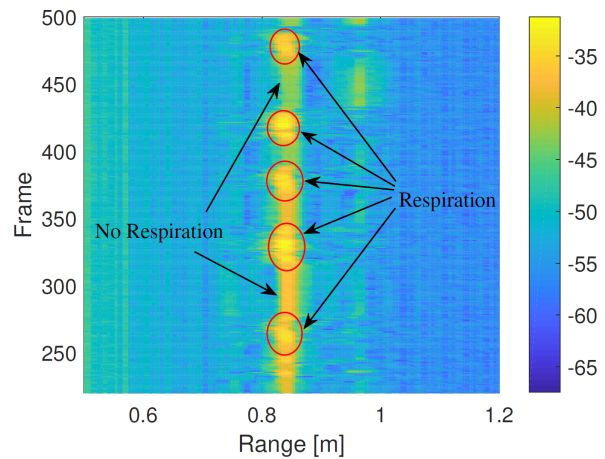

Fig. 2: Range waterfall and high power as respiration instants.

Two main effects can be observed in the filtered rang profile peaks. One effect is the target range and the other effect is the chest motion in both power and range dimensions. To remove the range aspect from our next computations, CFAR is used to filter out the target peaks of existing human subjects and analysis is done on each peak separately. Time domain reconstruction is applied using Inverse Fast Fourier Transform (IFFT) on the filtered peaks. The reconstructed time domain signal amplitude is highly correlated with the chest movements of the subject. Thresholding is applied to remove high random motions and false peaks not related to required activity. The reconstructed signals in the time domain yield consecutive impulses that represent an ECG-similar behaviour.

Envelope detection is then analysed on the output signals in the time domain to represent the respiration activity. The power-based technique shows clearly the detected irregular respiration activity as expected. Respiration is explained as a non-perfect periodic activity that differs through time and from one person to another. The respiration activity also changes with respect to the emotional level of the human [12]. Accordingly, the respiration pattern detection is very crucial for monitoring applications. The proposed technique in this paper achieves accurate respiration monitoring on real-time basis that represents the chest micro-motions as shown in Fig. 3

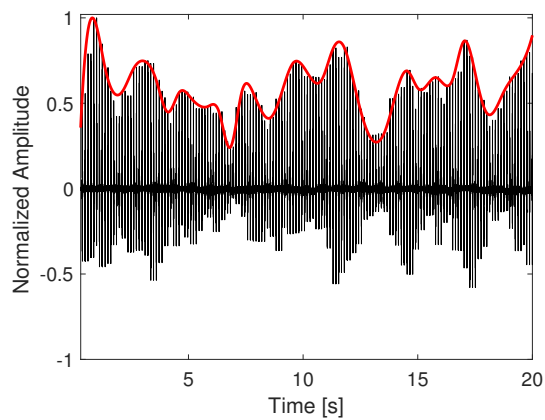

Fig. 3: Extracted Respiration activity.

## III. EXPERIMENTAL SETUP

### A. Distance and Aspect Angle Variations

Different aspects are taken into consideration in the applications designed for radar human interactions e.g. using radar for pedestrian recognition. The human different clothing is expected to influence the scattering of the human RCS and cause measurement problems for the real-time measurements. Accordingly, the effect of different materials of the human clothing and the accessories are studied when using the radar for human activity classification [13]. To study this effect on the presented technique, a human was tested wearing heavy winter clothes as shown in Fig. 4. Moreover, the respiration monitoring capability is tested for the maximum detectable distance with heavy clothing conditions.

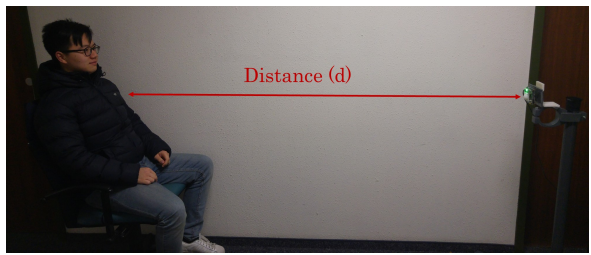

Fig. 4: Respiration monitoring with heavy winter clothes.

The operation idea of the radar is estimating the moving target velocity based on the analyzed Doppler frequency shift due to this motion. The perfect condition for the radar to estimate the target velocity is when both the target and the radar are aligned on the same line of sight with aspect angle ($\phi = 0°$). The availability of an aspect angle causes fading of the estimated target radial velocity by the radar [2]. Thus, respiration monitoring based on the induced frequency shifts in the radar received signals only is not sufficient as this effect will be relatively small due to the chest micro-motions. The experimental setup included measurements at different aspect angles from the radar to study the effect of this phenomena on the presented power-based analysis as shown in Fig. 5

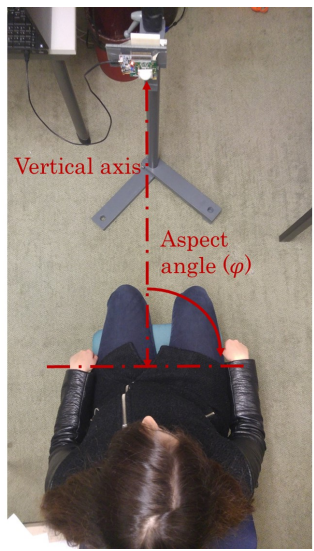

Fig. 5: Respiration monitoring for different aspect angles.

### B. Experimented Radar Modules

The power based respiration monitoring presented technique was tested using two different FMCW radar modules. The first module shown in Fig. 6a is a compact sensor developed by Silicon Radar [14]. The radar has a carrier frequency of $f_c = 120\,\text{GHz}$ and a bandwidth $B = 6\,\text{GHz}$. Then, the range resolution can be estimated based on relation in Eq. 1 given in [15] as $R_{res} = 2.5\,\text{cm}$, where $c$ is the speed of light. The output power of the module is limited; however, a maximum detectable range can reach up to 4 m. The Silicon radar has a beam aperture of only 3°, which can influence the aspect angle tests.

$$R_{res} = \frac{c}{2B} \qquad (1)$$

The proposed algorithm was also tested on another high end radar shown in Fig 6b developed by [17]. The radar works on a carrier frequency of $f_c = 94\,\text{GHz}$ and an extendable bandwidth of $B = 14\,\text{GHz}$. Then, the range resolution can be estimated based on Eq. 1 as $R_{res} \approx 1\,\text{cm}$. This module is working with a low noise amplifier (LNA) GaAs which enables better Signal to Noise Ratio (SNR) than the 120 GHz module. Moreover, the module gain is tunable. The maximum detectable range of this radar can reach a distance of 10 m. This radar beam aperture is about 11°.

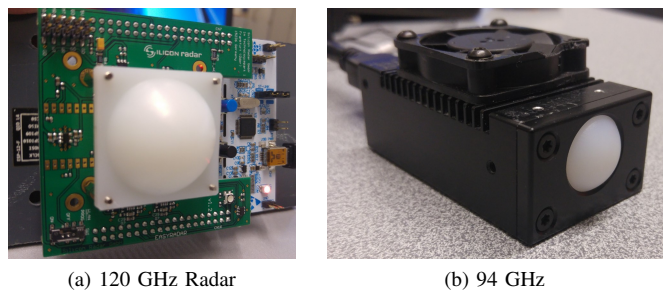

(a) 120 GHz Radar      (b) 94 GHz

Fig. 6: Experimented radars for respiration monitoring.

## IV. RESULTS

The experiments mentioned in the previous section was tested on different human subjects to extract respiration data. Since the extracted respiration signal can be detected in real-time during the respiration activity of the target, then no particular metric as respiration rate is needed to monitor the algorithm capability. The first experiment included subjects wearing heavy clothes standing in the radar line of sight and the respiration signal is extracted while varying the distance between the subject and radar. In the second experiment, the subject was sitting at a distance of 2 m in front of the radar and the radar was aligned with the chest at the same height. The following test was based on changing the subject orientation over the radar vertical axis as shown in Fig. 5. At each measurement the aspect angle $\phi$ is collected and the respiration signal is monitored. Finally, the maximum angle where no respiration signal can be furtherly visualized is recorded. Both tests were applied using both described radar modules.

The results of the experiments based on each proposed module is summarized in Table I. The 94 GHz module showed better performance than the 120 GHz in terms of both maximum aspect angle and distance where a respiration signal can still be monitored. This can be justified due to the higher aperture and output power in the 94 GHz module. However, the range resolution aspect did not show much influence in the attained results.

TABLE I: Comparison of different radar modules.

| Module/Experiment | 120 GHz Module | 94 GHz Module |
|---|---|---|
| Maximum distance | 2 m | 4.7 m |
| Maximum aspect angle ($\phi$) | 20° | 30° |

## V. CONCLUSION

This paper presents a technique for non-contact human respiration monitoring on real-time basis. The presented technique is based on using FMCW radar for its capability to jointly estimate both the target range and velocity. The analysis is based on observing the changes in the received power within the consecutive processed range profiles by the FMCW radar. Different prospectives were taken into consideration to evaluate the feasibility of the presented technique for real life conditions. First, the test is held on a human wearing winter heavy clothes to analyze the scattering effect of multi-layer clothing condition on the radar signal and the detected chest micro-motions. Second, tests were held while the human is sitting aligned with different aspect angles with the radar line of sight. Third, the maximum range that can yield extracted respiration monitoring is tested. The tests were made on two different radar modules with different specifications to validate the proposed power-based technique. The 120 GHz can monitor respiration till a distance of 2 m and an aspect angle of 20°. The 94 GHz module showed a very good performance and it was able to monitor the respiration signal at distances over 4 m and aspect angles of 30°. These results support the given specifications as the 94 GHz module has a higher SNR, gain and aperture.